\documentclass[12pt]{article}
\usepackage[totalheight = 23cm, totalwidth = 17cm]{geometry}

\newcommand{\be}{\begin{equation}}
\newcommand{\ee}{\end{equation}}
\newcommand{\bea}{\begin{eqnarray}}
\newcommand{\eea}{\end{eqnarray}}

\newcommand{\nn}{\nonumber\\}

\usepackage{graphicx}

\begin{document}

\begin{center}

{\bf{\Large An Alternative to Exact Renormalization 

and Cosmological Solutions in String Theory}

\vspace{.3cm}

Jean Alexandre} and {\bf Nick~E.~Mavromatos,}

\vspace{.2cm}

{\it Department of Physics, King's College London,

The Strand, London WC2R 2LS, U.K.}

\vspace{2cm}

{\bf Abstract}

\end{center}

\vspace{.3cm}

In this work we review the application of a functional method, serving
as an alternative to the Wilsonian Exact Renormalization approach, 
to stringy bosonic $\sigma$-models with metric and dilaton backgrounds
on a spherical world sheet \cite{AEM1}. We derive an exact evolution equation 
for the dilaton with the amplitude of quantum fluctuations, driven by the
kinetic term of the two-dimensional world-sheet theory. 
The linear dilaton conformal field theory, 
corresponding to a linearly (in cosmic Einstein-frame time) expanding 
Universe, appears 
as a trivial fixed point of this equation.
With the help of 
conformal-invariance conditions, we find a logarithmic 
dilaton as another, exact and non trivial, fixed-point 
solution, and determine the corresponding target-space 
physical metric. 
Cosmological implications of our solutions are briefly discussed, 
in particular the transition (exit) from the expanding Universe
of the linear dilaton to the Minkowski vacuum, corresponding to 
the non-trivial fixed point of our generalised flow. 
This novel renormalization-group method may therefore offer new insights
into exact properties of string theories of physical significance.  

\vspace{0.5cm}

{\it Talk presented in Demokritos National Research Centre (Athens, Greece),
in June 2005}

\vspace{1cm}

\section{Introduction}

Infinities in quantum field theory have been well understood today,
and in fact acquire an important physical significance, 
linked with the phase structure of the theory. 
They can be ``absorbed'' into appropriately renormalized 
operators and coupling constants, which then ``run'' 
with the scale, in a way characteristic of the theory or, better, 
class of theories.
The Renormalization Group (RG)~\cite{wilson} is therefore 
a very powerful method, which 
provides insight into the scale dependence of the various coupling 
constants of a quantum field theory, and through this,  
important information on the underlying phase structure.
In this way several important ``universality'' properties 
of physical theories, that is common behaviour 
of apparently different physical systems, have been understood   
by means of specific perturbations of ``fixed-point'' (scale invariant) 
theories. 

The so-called ``exact'' renormalization program~\cite{exact}, 
with the set up 
of appropriate flow equations for the coupling constants, 
helps compute the effective action of the theory in the limit 
where the cutoff goes to zero. In order to avoid the introduction of a running world sheet cut off, 
we present here an alternative approach which 
has been proposed in \cite{scalar}. The method dealt with the  
computation of the effective action of a theory  by introducing a parameter, 
which controls quantum fluctuations of the dynamical variables rather than
their Fourier modes, and looking at the 
corresponding evolution of the effective action, 
constructed as usual by means of 
a Legendre transform of the connected-graph generator functional.
This exact functional method was extended 
successfully to incorporate $QED$~\cite{QED}.

The idea is to control quantum fluctuations of a 
system by varying the bare mass of a theory, for a {\it fixed} cut off (or any other regulator).
When this mass is large, quantum fluctuations 
are frozen and the system is classical. As this bare mass 
decreases, quantum fluctuations gradually appear and the system gets dressed. 
The interesting point is that an exact evolution 
equation for the effective action
with the bare mass can be obtained, and it was shown that
the flows in this bare mass are equivalent 
to the usual renormalization flows at one loop.
Beyond one loop, the non perturbative aspect of this method 
leads to new insights into renormalization flows. A review
of this method, as well as applications to different models, 
can be found in \cite{reviewJanos}.

The Exact renormalization group method lies at the heart of 
the first quantization of string theory~\cite{polchinski}. Indeed, the latter 
is formulated as a two-dimensional world-sheet quantum field theory,
propagating in various target-space background fields, which from a 
two-dimensional viewpoint appear as (an infinite set of) 
``coupling constants'' $g^i$, with the index $i$ running in both
field-theory-species space and target space time,  
e.g. $g^i = \{ G_{\mu\nu}(y), 
\Phi (y) \dots \}$, with $y$ coordinates of a 
$D$-dimensional target-space time,  
$G_{\mu\nu}$ a metric tensor, $\Phi$ a dilaton field \emph{etc}.
In conventional RG approaches, 
the world-sheet infinities of the underlying (renormalizable) two-dimensional 
quantum field theory are absorbed in appropriately renormalized
background fields $g^i_R$, characterized by some $\beta$-functions 
$\beta^i = dg^i_R/d{\rm ln}(L/a)^2$, with ${\rm ln}(L/a)^2$ a 
Wilsonian world-sheet renormalization scale, with $L$ the size of the 
world-sheet of the string, and $a$ an appropriate short-distance cutoff. 
Removal of the (length) cutoff $a \to 0$, therefore, corresponds to 
infinite-size world sheets. 

In general, as in any other renormalized theory, 
there are relevant, marginal and irrelevant 
world-sheet deformations, which perturb a fixed-point theory away from 
the local scale invariant (conformal) point. 
Consistency of the underlying string theory requires 
conformal (local scale, Weyl) invariance, which in turn implies 
certain conditions on the coupling constants, characterizing 
the fixed-point theory. These conditions may be summarized by the 
vanishing of the so-called Weyl-anomaly coefficients of the $\sigma$-model,
related to the $\beta^i$-functions by 
${\tilde \beta}^i = \beta^i + \delta g^i$, where $\delta g^i$ are appropriate
functionals having the form of target-space 
diffeomorphism variations of the 
corresponding couplings/fields $g^i$. In this way, the fixed points 
in the theory space of stringy $\sigma$-models are not simply given 
by the vanishing of the 
$\beta^i$, but by the vanishing of 
\begin{equation}\label{weylfixedpoint} 
 {\tilde \beta}^i = \beta^i + \delta g^i = 0
\end{equation}
The extra terms $\delta g^i_R$ express the underlying 
target-space diffeomorphism invariance of the string theory,
whose couplings $g^i_R$ are fields in target space. The reader should 
notice that this latter property 
is not a feature shared by the coupling constants of ordinary 
field theories. The physical importance of (\ref{weylfixedpoint}) lies, 
of course, on the fact that it determines the consistent 
space-time backgrounds of strings, thereby providing important 
insight into the low-energy physics phenomenology~\cite{polchinski}. 

An important remark is in order at this point. 
In a perturbative $\alpha '$ (Regge slope) expansion, 
the leading order graviton $\beta$-functions contains 
the Ricci tensor of the string. Ignoring non-constant dilatons for the moment, 
this would imply that a consistent background (\ref{weylfixedpoint}) 
in strings is the Ricci flat space time, which is a vacuum solution of
a target-space (Einstein) effective theory.
In the presence of dilatons, the effective action becomes more complicated, 
of the form of a Brans-Dicke scalar/gravity theory~\cite{polchinski}. 

A non-trivial time dependence of the dilaton, 
in fact linear time dependence, has been shown in \cite{ABEN}
to be possible, satisfying the conditions (\ref{weylfixedpoint}), 
leading to string theory backgrounds with cosmological significance. 
The resulting space-time metrics where actually found to describe
Universes with a linearly expanding (or contracting) 
scale factor, in the so-called Einstein cosmic frame. 
The latter is a
system of target-space-time coordinates and field configurations, in which 
the metric field was redefined by appropriate exponential 
dilaton factors so as to ensure 
a canonical form for the Einstein curvature
term in the effective action, with no Brans-Dicke scale factors in front.
The cosmic time on the other hand, was defined 
in such a way that the metric in the 
Einstein frame had the standard Robertson-Walker form
\be
ds^2 = g_{\mu\nu}^E dx^\mu dx^\nu = -dt_E^2 + a^2(t_E)d{\vec x}^2~, 
\qquad {\vec x} = \{ x^1, \dots x^{D-1}\}
\label{RW}
\ee
For future use we remind the reader that 
the Einstein-frame metric 
$g_{\mu\nu}^E$ is related to the metric $G_{\mu\nu}$ 
of the original $\sigma$-model ($\sigma$-model-frame metric)
by~\cite{ABEN} 
\be
g_{\mu\nu} = e^{-4\Phi/(D-2)}G_{\mu\nu}
\label{smodeinst}
\ee
and the time $t_E$ 
is related to the time coordinate $X^0$ of 
the $\sigma$-model by: 
\be
dt_E = e^{-2\Phi/(D-2)} dX^0
\label{einsttime}
\ee
The appropriate dilaton configuration leading to 
(\ref{RW}) was linear in $X^0$:
\be\label{linearintro}
\Phi (X^0) =-QX^0~, ~~~~~~Q^2 = \frac{D-26}{3}
\ee
while the $\sigma$-model metric $G_{\mu\nu}= \eta_{\mu\nu}$ was 
Minkowski flat, corresponding, in the Einstein frame, to a
Robertson-Walker metric with a scale factor depending linearly  
on the Einstein time-$t_E$, which can be 
expanding or contracting, depending on the algebraic sign of $Q$. 
In fact, the solution of \cite{ABEN}  allowed
for a consistent formulation of strings in other than $D=26$ dimensions,
and it was the first example of a non-critical (supercritical)
string~\cite{ddk}, with the time $X^0$ playing the r\^ole 
of the Liouville mode, which has time-like signature.

Liouville strings correspond to $\sigma$-models which are perturbed by
deformations away from their conformal points in string theory space
$\{ g^i \}$. The conformal invariance is restored, provided the 
theory is ``dressed'' by the Liouville mode, $\varphi$, which appears as
an extra dynamical field in the $\sigma$-model, coupled to the 
world-sheet curvature term in the action, hence contributing to the
dilaton field, which thereby acquires special significance~\cite{ddk,ABEN}.
If the non-critical stringy $\sigma$-model 
has a central charge 
deficit, i.e. the central charge of the theory is less than that necessary for 
the $\sigma$-model to be critical, then the Liouville mode has space-like 
signature, in the sense that the kinetic term of the Liouville field in 
the two-dimensional action appears with the signature 
of the kinetic term of the target spatial coordinate fields in 
a stringy $\sigma$-model. On the other hand, if there is a central
charge surplus (supercritical string~\cite{ABEN}), the Liouville
field has time-like signature, and may play the r\^ole of the target 
time~\cite{ABEN,emn}. 

The restoration of conformal invariance by the presence of this 
extra $\sigma$-model field means that a non-critical string in D target-space 
dimensions can become a critical one in (D+1) dimensions.
However, in \cite{emn} an alternative to the traditional 
Liouville dressing procedure 
has been suggested, according to which
the dimensionality of the target space of the non-critical 
string remains the same as that before the dressing.
This can be achieved as follows: one starts with a non-critical $\sigma$-model 
formulated on a D-dimensional target space time $(X^0, {\vec x})$.
The initial non-criticality may be due to catastrophic cosmic events,
for instance the collision of two brane worlds~\cite{gravanis}.
The Liouville dressing procedure initially leads to a $(D+1)$-dimensional
target space, but, eventually, 
dynamical reasons~\cite{gravanis}, such as minimization
of appropriate effective potentials, impose the 
identification of the world-sheet zero mode of the Liouville 
field itself with (a function of) the target time $X^0$, thereby
keeping the target space $D$-dimensional.
Only certain backgrounds are consistent with such an identification,
but fortunately among those there are some with 
cosmological interest, including 
inflationary models~\cite{brany}, 
and in general expanding Universes in string theory~\cite{dgmpp},
which for large cosmic times asymptote to the linear-dilaton solutions of 
\cite{ABEN}. 

The Liouville dressing procedure may be viewed in some sense as a special 
renormalization of the two-dimensional world-sheet theory,
in which the RG scale is itself a dynamical field, responsible for the 
restoration of conformal invariance.  The zero mode of the Liouville
field may be related to the logarithm of the world-sheet area, which thus
provides the running scale. The restoration of conformal invariance 
implies the independence of the target-space dynamics from world-sheet 
physics. 

An exact RG approach to Liouville strings was 
applied in \cite{reuter},
where the standard quantum Liouville field theory was derived as a solution 
of a flow equation.
However, the two-dimensional theory suffers from the same inconsistencies
and ambiguities 
of the higher-dimensional analogue, mentioned in the beginning of the
section. For instance, a truncation of the space of action functionals
is necessary in the analysis of \cite{reuter}, which although consistent
with Weyl invariance, nevertheless is not unambiguous and moreover its
physical significance is not entirely clear. 

{}From such a point of view, it then seems necessary 
to explore the possibility of applying alternative approaches, 
such as the 
functional method of \cite{scalar}, to stringy $\sigma$-models. 
In our framework 
both critical and 
non critical $\sigma$-models 
appear formally equivalent, 
since, as mentioned above, a Liouville-dressed 
string is a critical one in one dimension
higher, or sometimes~\cite{emn} even in the same dimension. 
In particular, we shall be interested in 
finding novel  
cosmological backgrounds in string theory by following
such alternative ``renormalization'' methods.
As we shall see in this work, this novel method is closer in spirit to 
the Liouville approach of \cite{emn,brany}, 
in which although one starts with an independent
scale, the Liouville mode, eventually the latter
is identified with a function of the cosmic time. 
A similar thing will characterize our approach:
the control parameter, which plays the r\^ole of a running scale
in the approach of \cite{scalar}, 
becomes a function of the time coordinate of 
the string after quantum dressing, and this 
will lead to interesting cosmological backgrounds of strings,
corresponding to fixed points of the alternative flow equations. 
As we shall demonstrate \cite{AEM1}, though, such backgrounds 
will necessarily characterize only non-critical dimension strings. 

Moreover, our new renormalization method will lead to 
results of physical significance 
that one could not obtain in conventional 
world-sheet renormalization-group approaches.
In particular, in the expanding Universe solution
of \cite{ABEN} it was not clear how  
the Universe 
exits from that phase into the static Minkowski space time,
which could serve as an equilibrium point.
As we shall show in this work, our new approach offers an
interesting solution to this puzzle. The Minkowski
space time corresponds to a non-trivial fixed point of the 
generalized flow of our control parameter, with a non-trivial
dilaton field, while the linear dilaton solution,
with a linearly (in Einstein-frame cosmic time) expanding 
scale factor, is simply the trivial fixed point of the flow.
Thus, a flow of the control parameter from the trivial 
to the non-trivial fixed point, standard in 
any renormalization approach, provides such an exit from the 
expanding Universe to the static Minkowski phase, 
in view of the control parameter becoming a function of cosmic time
after quantum dressing.

However, from a technical point of view it should be stressed that,
if one wishes to apply the alternative method of 
\cite{scalar,QED} to the world-sheet field theory 
of a stringy $\sigma$-model, one needs a modification. 
In the case of bosonic or supersymmetric strings propagating in 
the background of the 
massless string-multiplet, which we shall be dealing with 
in this work, i.e. in the absence of tachyons, 
no mass term is present in the two-dimensional world-sheet action. 
Nevertheless, the same 
method as in \cite{scalar} can be used, but now  
one should be looking at the evolution of the quantum theory with the 
amplitude of the kinetic term, 
controlled by a dimensionless parameter $\lambda$.

The structure of the article is as follows: the 
evolution equation for the 
dilaton with respect to the control 
parameter $\lambda$ is derived in Section 2.
We study the fixed point 
solutions of this exact equation, 
taking into account also the Weyl invariance constraints
of the $\sigma$-model.
We first demonstrate, as a consistency check, 
that the linear dilaton configuration 
of \cite{ABEN} is an exact solution
of our evolution equation, corresponding to a trivial 
fixed point. 
Then we proceed, in section 3,  
to find another exact solution, 
pertaining to a non-trivial fixed point of the evolution
with $\lambda$. This solution is characterized, in four large 
(uncompactified) dimensions,
by a dilaton configuration, 
which is logarithmic in the $\sigma$-model-frame time, $X^0$,  
and a target-space metric, which, in the Einstein physical frame, 
turns out to be the (static) Minkowski space time.
Cosmological implications 
of the non trivial solution, in particular
its role as providing the exit phase of 
the linearly expanding Universe of \cite{ABEN},
are discussed briefly in Section 4, emphasizing the fact that only
non-critical-dimension string models can admit such 
time-dependent vacua. The non trivial solution in dimensions
other than four can accommodate accelerating Universes
without a cosmic horizon. 
Finally, in section 5 we present our Conclusions and Outlook.

\section{Exact evolution equation}

We consider a spherical world sheet with a curvature scalar $R$. 
The bare action 
of the Bosonic $\sigma$-model in Graviton and Dilaton  
backgrounds reads~\cite{polchinski}
\bea
S&=&\frac{1}{4\pi}\int d^2\xi\sqrt{\gamma}\left\{\gamma^{ab}\lambda\eta_{\mu\nu}\partial_a X^\mu\partial_b X^\nu
+R\phi_B(X^0)\right\},
\eea
where $\eta_{\mu\nu}$ is the flat Minkowski target-space metric. 
The parameter $\lambda$ plays the role of $1/\alpha^{'}$ and controls the amplitude of 
the kinetic term: for $\lambda>>1$, the latter
dominates over the bare dilaton $\phi_B$ term, and 
the theory is classical. As $\lambda$ decreases, the effects
of the interactions in $\phi_B$ gradually appear 
and the quantum theory settles in.

We note the following points:
\begin{itemize}
\item 
The variables $X^k$, $k\ne 0$, do not play 
a dynamical role in our study since we consider a dilaton
depending on $X^0$ only. This is because we are interested, in the 
spirit of \cite{ABEN}, in constructing cosmological backgrounds only. 
\item
The expression for the 
classical dilaton $\phi_B$ is not explicitly given here, 
but it should in principle
contain interactions (at least cubic in $X^0$) so 
as to generate quantum fluctuations.
\end{itemize}

\vspace{.5cm}

The quantum theory is described by the effective action $\Gamma$, i.e. 
the proper-graph-generating functional, for which 
the exact evolution equation with respect to the control parameter $\lambda$, which reads \cite{AEM1}: 
\bea\label{evolG}
\dot\Gamma_\lambda&=&
\frac{1}{4\pi}\int d^2\xi\sqrt{\gamma}\gamma^{ab}\eta_{\mu\nu}\partial_a X^\mu\partial_b X^\nu\\
&&+\frac{\eta_{\mu\nu}}{4\pi}\mbox{Tr}\left\{\gamma^{ab}
\frac{\partial}{\partial\xi^a}\frac{\partial}{\partial\zeta^b}
\left(\frac{\delta^2\Gamma_\lambda}{\delta X^\mu(\zeta)\delta X^\nu(\xi)}\right)^{-1}\right\},\nonumber
\eea
where a dot over a letter denotes a derivative with respect to $\lambda$.
In eq.(\ref{evolG}), the symbol of the trace 
contains the quantum corrections to $S$. In 
order to obtain physical information on the system,
eq.(\ref{evolG}) should  
in principle be integrated from $\lambda=\infty$ to $\lambda=1$, 
which is the appropriate regime of the
full quantum theory. In the present article, though, 
we shall be dealing only with fixed-point
solutions of eq.(\ref{evolG}), and hence 
we shall not follow the $\lambda$-dependence. This is left for a future 
work.

We now derive the evolution equation for the quantum dilaton with $\lambda$.
For this one must have knowledge of the functional dependence
of $\Gamma$ on the quantum fields. This can be achieved by 
means of a gradient-expansion approximation, which assumes that,
for any value of $\lambda$, $\Gamma$ takes the form:
\bea\label{gradexp}
\Gamma&=&\frac{1}{4\pi}
\int d^2\xi\sqrt{\gamma}\left\{\gamma^{ab}\kappa_\lambda(X^0)\partial_a X^0\partial_b X^0\right.\nn
&&~~~~~~~~~~~\left.+\gamma^{ab}\tau_\lambda(X^0)\partial_a X^k\partial_b X^k+R\phi_\lambda(X^0)\right\},
\eea
where $\kappa,\tau$ are $\lambda$-dependent functions of $X^0$. The latter 
are different for the time ($X^0$) and space ($X^k$) coordinates, since the respective 
quantum fluctuations are different.

The approximation (\ref{gradexp}), when plugged in the exact
evolution equation (\ref{evolG}), leads to, in the limit of a flat world-sheet metric \cite{AEM1}:
\be\label{evolphi}
\dot\phi=-\frac{\Lambda^2}{2R^{(2)}}\left(\frac{1}{\kappa}+\frac{D-1}{\tau}\right)+
\frac{\phi^{''}}{4\kappa^2}\ln\left(1+\frac{2\Lambda^2\kappa}{R^{(2)}\phi^{''}}\right)\nonumber,
\ee
where $\Lambda$ is the world 
sheet UV cut off and a prime denotes a derivative with respect to $X^0$.

For consistency of eq.(\ref{evolphi}), we can check that the linear dilaton
configuration (\ref{linear}) of \cite{ABEN}, corresponding to a
Minkowski flat $\sigma$-model-frame metric, or, equivalently, to a
power-law expanding Robertson-Walker Universe in the
Einstein-cosmic-time-$t_E$ frame,
is an exact solution.
After the redefinition $X^j\to\sqrt{D-1}X^j$ of the space coordinates of the string,
one can see that the evolution equation (\ref{evolphi}) is satisfied by the well-known
flat metric/linear dilaton configuration~\cite{ABEN}:
\be\label{linear}
\kappa=1=-\tau,~~~~~~~~\phi(X^0)=QX^0,
\ee
which shows that the latter solution is exactly marginal with respect to the flows in $\lambda$.

\section{Non trivial fixed-point solution}

Besdies the expected flat metric/linear dilaton configuration, the evolution equation 
(\ref{evolphi}) has another non-trivial solution.

We consider a configuration with $\kappa(X^0)=F\phi^{''}(X^0)$, where $F$ is a constant. For such a
configuration, the evolution equation (\ref{evolphi}) reads
\be
\kappa\dot\phi=-\frac{\Lambda^2}{2R^{(2)}}\left(1+(D-1)\frac{\kappa}{\tau}\right)
+\frac{1}{4F}\ln\left(1+\frac{2\Lambda^2F}{R^{(2)}}\right),
\ee
and one can see that it is possible to have a $\lambda$-independent solution: $\dot\phi=0$, if
\be
\label{condition}
\frac{\kappa}{\tau}=-\frac{1}{D-1}+\frac{R^{(2)}}{2(D-1)\Lambda^2F}
\ln\left(1+\frac{2\Lambda^2F}{R^{(2)}}\right)
=-c^2,
\ee
where we have taken the negative sign corresponding to Minkowski signature, as is appropriate for large cut-off $\Lambda$,
in which case the ratio $\kappa/\tau$ is necessarily negative, and $c$ is a positive constant.
After the redefinition $X^j\to cX^j$ of the space coordinates of the string,
the condition (\ref{condition}) shows that the target-space metric is conformally flat, and the
non-trivial $\lambda$-independent solution of eq.(\ref{evolphi}) is such that
\be\label{nontrivial}
g_{\mu\nu}(X^0)\propto\phi^{''}(X^0)\eta_{\mu\nu}.
\ee
In the stringy $\sigma$-model framework, any equilibrium
solution, i.e., one that satisfies the equations of motion of a target-space effective action,
must also satisfy the conformal-invariance
conditions~\footnote{Strictly, the Weyl-invariance conditions,
that take into account target-space diffeomorphisms~\cite{metsaev}.} on the world-sheet.
{\it A priori}, it is not clear that the configuration (\ref{nontrivial}),(\ref{condition})
satisfies these conditions. However, in the next Section we
display a more precise functional dependence for $\phi$, which we conjecture satisfies the
the Weyl invariance conditions. This configuration has the form:
\begin{equation}
ds^2 =\frac{\alpha ' A}{(X^0)^2}\left(-(dX^0)^2 + dX_i dX_i\right), \quad
\phi = \phi_0 {\rm ln}X^0
\label{specific} 
\end{equation}
where $\phi_0$ and $A$ are to be determined.

This is still a conjecture, since the new fixed-point configuration (\ref{specific}) of the $\lambda$-
low is non perturbative in $\alpha '$, so the corresponding Weyl anomaly coefficients
are not known in a closed form. We base our conjecture that it is indeed conformally invariant
on a heuristic inductive argument that there exists in principle a world-sheet renormalization scheme, 
reached from the standard $\sigma$-model scheme~\cite{metsaev} by certain field redefinitions, in which the
Weyl anomaly coefficients vanish. We now demonstrate this explicitly to order $\alpha '$ and then use 
inductive arguments to argue that this is true to all orders in $\alpha '$.

To first order in $\alpha^{'}$, the beta functions for the bosonic world-sheet
$\sigma$-model theory in graviton and dilaton backgrounds are \cite{metsaev}:
\bea\label{weylconditions}
\beta_{\mu\nu}^g&=&R_{\mu\nu}+2\nabla_\mu\nabla_\nu\phi
+\frac{\alpha^{'}}{2}R_{\mu\lambda\rho\sigma}R_\nu^{~~\lambda\rho\sigma}+{\cal O}(\alpha^{'})^2,\nn
\beta^\phi&=&\frac{D-26}{6\alpha^{'}}-\frac{1}{2}\nabla^2\phi+\partial^\rho\phi\partial_\rho\phi
+\frac{\alpha^{'}}{16}R_{\mu\rho\nu\sigma}R^{\mu\rho\nu\sigma}+{\cal O}(\alpha^{'})^2.
\eea
We consider first the tree-level beta functions for a configuration satisfying the
condition $\kappa(X^0)\propto\phi^{''}(X^0)$, with the power-law dependence
\bea\label{ansatz}
\phi^{'}(X^0)&=&\phi_0(X^0)^n,\nn
\kappa(X^0)&=&\kappa_0(X^0)^{n-1}.
\eea
We observe that such a ansatz does not satisfy perturbative Weyl invariance conditions,
but the important point is that, for $n=-1$, all the other orders in $\alpha^{'}$ are,
as functions of $X^0$, homogenous to the tree level: we can write
\bea
\beta^g_{00}&=&\frac{1}{(X^0)^2}\sum_{m=0}^\infty \xi_m\left(\frac{\alpha^{'}}{\kappa_0}\right)^m,\nn
\beta^g_{jk}&=&\frac{\delta_{jk}}{(X^0)^2}\sum_{m=0}^\infty \zeta_m\left(\frac{\alpha^{'}}{\kappa_0}\right)^m,\nn
\beta^\phi&=&\frac{1}{\alpha^{'}}\sum_{m=0}^\infty \eta_m\left(\frac{\alpha^{'}}{\kappa_0}\right)^m,
\eea
where $\xi_m,\zeta_m,\eta_m$ are $\alpha^{'}$-independent coefficients. As a consequence,
for $\kappa_0$ of the same order as $\alpha^{'}$, the expansion of the beta functions in $\alpha^{'}$
is no longer valid.

The next step is to argue that the configuration
\bea\label{config}
\phi(X^0)&=&\phi_0\ln(X^0),\nn
\kappa(X^0)&=&\frac{\alpha^{'}A}{(X^0)^2},
\eea
where we write $\kappa_0=\alpha^{'}A$,
may satisfy conformal invariance at a non-perturbative level.
We exploit the fact, well-known in string theory, that at higher orders in $\alpha^{'}$ the beta functions  are not fixed uniquely, but can be changed by making local
field redefinitions~\cite{metsaev}: $g_{\mu\nu}\to\tilde g_{\mu\nu}$ and $\phi\to\tilde\phi$,
which leave the (perturbative) string S-matrix amplitudes invariant.
This possibility of field redefinition enables us to maintain conformal invariance to
all orders in $\alpha^{'}$.
 
We illustrate this possibility with an explicit calculation to first order in $\alpha^{'}$.
In our case, since we keep the target-space metric
conformally flat, the redefinition of the metric must be such that $\tilde g_{\mu\nu}$
is proportional to $g_{\mu\nu}$, and thus we can consider the following redefinitions:
\bea\label{redef}
\tilde g_{\mu\nu}&=&g_{\mu\nu}+\alpha^{'}g_{\mu\nu}
\left(b_1R+b_2\partial^\rho\phi\partial_\rho\phi+b_3\nabla^2\phi\right),\nn
\tilde\phi&=&\phi+\alpha^{'}
\left(c_1R+c_2\partial^\rho\phi\partial_\rho\phi+c_3\nabla^2\phi\right),
\eea
where $b_1,b_2,b_3,c_1,c_2,c_3$ are constants and $R,\nabla$
corresponds to the metric $g_{\mu\nu}$.
In the case of the configuration (\ref{config}), we have (see Appendix B for details):
\bea\label{gtilde}
\tilde g_{\mu\nu}&=&g_{\mu\nu}+\frac{g_{\mu\nu}}{A}\Big(-b_1(D-1)^2+b_2\phi_0^2-b_3(D-1)\phi_0\Big)=(1+B)g_{\mu\nu},\nn
\tilde\phi&=&\phi+\frac{1}{A}\Big(-c_1(D-1)^2+c_2\phi_0^2-c_3(D-1)\phi_0\Big)=\phi+C,
\eea
where $B,C$ are constants linear in $b_1,b_2,b_3,c_1,c_2,c_3$.
Therefore, the redefinitions (\ref{redef}) consist of adding a constant to the dilaton and rescaling
the metric, thus not changing the functional dependence of the configuration (\ref{config}).

The new beta functions
$\beta_{\mu\nu}^g\to\tilde\beta_{\mu\nu}^g$ and $\beta^\phi
\to\tilde\beta^\phi$ are obtained via the appropriate Lie derivatives in theory
space~\cite{metsaev}, as appropriate to the vector nature of the
$\beta^i$ functions in this space:
\bea\label{betatildeg}
\tilde\beta_{\mu\nu}^g-\beta_{\mu\nu}^g
&=&\int(\tilde g_{\rho\sigma}-g_{\rho\sigma})\frac{\delta\beta_{\mu\nu}^g}{\delta g_{\rho\sigma}}
+\int(\tilde\phi-\phi)\frac{\delta\beta_{\mu\nu}^g}{\delta\phi}\nn
&&-\int\beta_{\rho\sigma}^g\frac{\delta(\tilde g_{\mu\nu}-g_{\mu\nu})}{\delta g_{\rho\sigma}}
-\int\beta^\phi\frac{\delta(\tilde g_{\mu\nu}-g_{\mu\nu})}{\delta\phi},
\eea
and
\bea\label{betatildephi}
\tilde\beta^\phi-\beta^\phi
&=&\int(\tilde g_{\rho\sigma}-g_{\rho\sigma})\frac{\delta\beta^\phi}{\delta g_{\rho\sigma}}
+\int(\tilde\phi-\phi)\frac{\delta\beta^\phi}{\delta\phi}\nn
&&-\int\beta_{\rho\sigma}^g\frac{\delta(\tilde\phi-\phi)}{\delta g_{\rho\sigma}}
-\int\beta^\phi\frac{\delta(\tilde\phi-\phi)}{\delta\phi}.
\eea
It is shown in \cite{AEM1}, at one loop, that the beta functions obtained after this change of string
parametrization can indeed be cancelled. 

This analysis takes into account only the first non-trivial order in $\alpha^{'}$, whereas all the
higher orders should also be taken into account. However, this first-order analysis
provides the basis for an inductive argument. If conformal invariance is satisfied at
order $n$ in $\alpha^{'}$, as in the first-order case worked out above, there are always
enough parameters in the redefinitions of the metric and dilaton at the next order,
leaving the string configuration unchanged, which enable the beta functions to vanish and hence
conformal invariance to be satisfied at the
next order $n+1$ in $\alpha^{'}$.
                                                                                                                             
We stress again that these arguments are only heuristic at this stage,
since the $\beta$-functions and Weyl anomaly coefficients are not exactly known
to all orders in $\alpha '$, and hence the world-sheet renormalization scheme
in which they vanish is abstract. We mention at this point that it is known from standard analyses of
$\sigma$ models~\cite{metsaev} that there is a formal scheme in which
the dilaton dependence is simply that of one $\sigma$-model
loop. For the graviton Weyl-anomaly coefficient this means that the dilaton
dependence has the form of a target-space diffeomorphism : $\nabla_\mu \partial_\nu \phi$.
This is {\it not} the scheme we use in this work, and thus the reader should bear in mind that
the background (\ref{specific}) proposed here
does not correspond to a standard conformal field theory that is
perturbative in $\alpha '$. This reflects the
non-perturbative nature of the novel renormalization-group method
employed in our approach.

\section{Cosmological implications}

We now examine the physical significance of the new non-trivial fixed-point solution (\ref{config}),
and discuss briefly its cosmological implications. This leads to a value of the constant $\phi_0$
that appears in the configuration (\ref{config}).
                                                                                                                             
The relation between the physical metric in the Einstein frame and the string metric is given
by~\cite{ABEN}
\bea
ds^2&=&dt^2-a^2(t)dx^kdx^k\nn
&=&\kappa(x^0)\exp\left\{-\frac{4\phi(x^0)}{D-2}\right\}\left(dx^0dx^0-dx^kdx^k\right),
\eea
where $a(t)$ is the scale factor of a spatially flat Robertson-Walker-Friedmann Universe,
and the $x^\mu$ are the zero modes of $X^\mu$. From the configuration (\ref{config}), we have
\be
\frac{dt}{dx^0}=\varepsilon\sqrt{|\kappa_0|} (x^0)^{-1-\frac{2\phi_0}{D-2}},
\ee
where $\varepsilon=\pm 1$, such that
\be\label{tx^0}
t=T+\sqrt{|\kappa_0|}\frac{(D-2)}{2|\phi_0|}(x^0)^{-\frac{2\phi_0}{D-2}},
\ee
where $T$ is a constant. We find then a power law for the evolution of the scale factor:
\be\label{scaleexpans}
a(t)=a_0|t-T|^{\frac{D-2}{2\phi_0}+1},
\ee
which is in general singular as $t \to T$.
                                                                                                                             
In order to have a Minkowski target space, one needs
$D-2+2\phi_0=0$. As was discussed earlier,
the choices of $D$ and $\phi_0$ are free, and lead to the
determination of $\kappa_0$ (in a way which has not been determined yet).
As a consequence, for a given dimension $D$, it is always
possible to choose $\phi_0$ so that the target space is static and flat.
It may therefore find an application to the exit phase from the
linearly expanding Universe associated with the linear dilaton of~\cite{ABEN}.
                                                                                                                             
We note that, in terms of the Einstein time $t$, the dilaton can be written, up to a constant, as:
\be\label{logdil}
\phi=-\frac{D-2}{2}\ln|t-T|.
\ee
We observe that, like the scale factor (\ref{scaleexpans}), the dilaton has a
singularity as $t \to T$. It would be interesting to explore the applicability of
this configuration to primordial cosmology. The sign of the expression (\ref{logdil}) for the
dilaton when $D > 2$ ensures that the string coupling is small at large times.

Finally, we mention for completeness that our solution (\ref{config}) 
(or (\ref{scaleexpans}),(\ref{logdil}) in the Einstein frame) should be compared
and contrasted with the (isotropic-case) solution of \cite{mueller},
which describes a D-dimensional Universe whose constant-time slices are (D-1)-dimensional tori
with time-dependent `rolling'' radii, whose flow is attributed to the existence of a rolling dilaton field.
First, it should be remarked that the work of~\cite{mueller} is perturbative in $\alpha^{'}$, unlike
our construction. Moreover, our main point in this work has been to associate the non-trivial fixed
point solution (\ref{config}) to a marginal configuration of flow with respect to a novel control
parameter. In this way, we have provided arguments
for the r\^ole of the solution, especially in the Minkowski case, as providing an exit phase from
a linearly-expanding Universe. None of these aspects apply to the work of \cite{mueller}.

\section{Conclusions}

We have proposed here a new non-perturbative renormalization-group technique
for the bosonic string, based on a functional method for controlling the quantum fluctuations,
whose magnitudes are scaled by the value of $\alpha^{'}$. Using this technique, we have exhibited a
new, non-perturbative time-dependent background solution. Using the
field redefinition ambiguities of the target-space effective action, which leave the string S-matrix invariant,
we have demonstrated that this
solution is conformally invariant to ${\cal O}(\alpha^{'})$, and we have made a conjecture,
based on a heuristic and inductive argument,
that conformal invariance can be maintained to all orders in $\alpha^{'}$.
We stress once again that our work, which is based on the flow equations of a non-perturbative
Schwinger-Dyson-type effective action,
is different in spirit from other work in the literature~\cite{gubser}, where expressions for
the $\beta$ functions to all orders in $\alpha'$ are obtained in a large-N-treatments, where N is
the number of target-space dimensions.
                                                                                                                             
This new non-perturbative time-dependent background solution has related singularities in both
the metric scale factor and the dilaton value at a specific value of the time in the Einstein frame.
A full exploration of the possible cosmological applications of this solution lies beyond the
scope of this paper, but we do note two interesting possibilities. One is that the temporal
singularity might be relevant for primordial cosmology, i.e., the beginning of the Big Bang. The
second possible application could be to describe the exit phase from the
linearly expanding Universe associated with the linear dilaton of~\cite{ABEN}.
                                                                                                                             
These are two phenomenological tasks for future work on this new non-perturbative
time-dependent background solution. It is also desirable to explore in more detail the
formal underpinnings of the solution. In particular, it is necessary to improve on our
heuristic inductive argument that its conformal invariance may be maintained to all orders
in $\alpha^{'}$. We also note that the non-perturbative renormalization-group technique
proposed here may have applications to other aspects of string theory.

There may be other non-trivial fixed point solutions
of our flow equations, especially if more backgrounds,
such as antisymmetric tensors and fields from the matter multiplet of
strings, are included.
In the spirit of what was discussed in this paper,
such solutions may provide explanations on the phase
transition of various stages of the string Universe, from grateful
exit from inflation and reheating, to issues of possible
pre-Big-Bang cosmologies\cite{stringycosmol},
which could not be
explained within the context of conventional
world-sheet renormalization approaches to the low-energy limit of strings.

Before closing we would like to remark
that it is also possible to apply the same
method to a string with a tachyonic
background. Such backgrounds may play
an important r\^ole in early brane-Universe cosmology~\cite{dgmpp,sen},
since they can provide the initial cosmological
instability, decoupling relatively quickly from the
spectrum. The tachyonic background involves
a mass term for the world-sheet $\sigma$-model fields,
and a usual Wilsonian Exact Renormalization method has been
applied to this case \cite{dwyer}, where
the equation of motion of the quantum tachyon is found
as a consequence of the independence
of the theory from the world sheet cut off.
This situation can be studied
with the present method, where quantum fluctuations of
the tachyon are controlled by the mass parameter
in the world-sheet theory.
The resulting evolution equation of the quantum
theory would look like eq.~(\ref{evolG}), but without
derivatives with respect to the world sheet
coordinates inside the trace. A suitable
derivative expansion for the effective action
(together with the Weyl invariance conditions),
plugged in this evolution equation, would then
give coupled renormalization group equations
for the tachyon and metric backgrounds.
As in the case of the dilaton, studied
in the present paper, a cut-off free
evolution equation would then be obtained,
enabling therefore one to study the quantum theory
by means of renormalized physical quantities,
in a way independent of unphysical cut off scales.
We shall return to this issue in a future publication.

\end{document}